\documentclass[aps,prl,twocolumn,groupedaddress]{revtex4}

\usepackage{graphicx}
\begin{document}

\newcommand{\esp}{\epsilon_{sp}}
\newcommand{\ed}{\epsilon_{d}}
\newcommand{\ec}{\epsilon_{c}}
\newcommand{\e}{\epsilon}

\title{Anomalous dispersion of optical phonons  at the neutral-ionic transition:\\
 Evidence from  diffuse X-ray scattering }

\author{Gabriele D'Avino}
\affiliation{Dip. Chimica GIAF, Parma University, \& INSTM UdR Parma,
  43100 Parma, Italy}
\author{Alberto Girlando}
\affiliation{Dip. Chimica GIAF, Parma University, \& INSTM UdR Parma,
  43100 Parma, Italy}

\author{Anna Painelli}
\email[]{anna.painelli@unipr.it}
\homepage[]{http://continfo.chim.unipr.it/mmaa/}
\affiliation{Dip. Chimica GIAF, Parma University, \& INSTM UdR Parma,
  43100 Parma, Italy}

\author{Marie-H\'el\`ene Lem\'ee-Cailleau}
\affiliation{Institut Laue-Langevin, Grenoble, France}

\author{Zolt\'an G. Soos}
\affiliation{Dept. Chemistry, Princeton University, Princeton, New
  Jersey 08544}


\date{\today}

\begin{abstract}
Diffuse X-ray  data for mixed stack organic 
charge-transfer crystals approaching the
neutral-ionic phase transition can be quantitatively explained as
due to the softening of the {\it optical} phonon
branch. The interpretation is fully consistent with vibrational spectra,
and underlines the importance of electron-phonon coupling in
low-dimensional systems with delocalized electrons.
\end{abstract}


\maketitle

The physics of low-dimensional materials is governed by the complex
interplay between electron-electron
(e-e) and electron-phonon (e-ph) interactions. The resulting
phenomenology is very rich and includes complex  phase
diagrams, competing phases, 
quantum phase transitions and multistability. 
Solitons and domain boundaries are intriguing low-energy excitations 
in 1-dimensional (1D) systems such as polyacetylene \cite{ssh} or
organic charge-transfer (CT) salts \cite{nag3}.
 The neutral-ionic phase transition (NIT) observed in CT 
crystals with mixed stacks \cite{ricememorial} offers unique opportunities for 
investigating this rich physics. NIT is an {\it electronic} quantum phase 
transition, driven by the  volume compression on cooling or under pressure, in 
which a stack of largely neutral molecules with regular spacing 
...D$^{\rho +}$A$^{\rho -}$D$^{\rho +}$A$^{\rho -}$... ($\rho <0.5$)
switches to a dimerized stack of more ionic molecules,  ...(D$^{\rho
+}$A$^{\rho -}$)(D$^{\rho +}$A$^{\rho -})$.... or  
 ...(A$^{\rho -}$D$^{\rho +}$)(A$^{\rho -}$D$^{\rho +})$ ($\rho
>0.5$). Dimerization is a {\it structural} or Peierls transition driven by e-ph
interactions. NIT with discontinuous changes in $\rho$ and in
the dimerization amplitude ($\delta$) are observed in some CT crystals,
while others show continuous variations of $\rho$ and $\delta$
\cite{ricememorial}. 

In addition to domain walls or topological solitons \cite{ssh} that 
separate regions with opposite dimerization, CT crystals with mixed
stacks have neutral-ionic domain walls (NIDW) that separate neutral and 
ionic regions along the 1D stack. Nagaosa \cite{nag3} introduced the
concept of NIDW and the corresponding excitations or domains,
 called lattice-relaxed 
exciton strings, LR-CT. These concepts were subsequently used to
discuss the anomalies observed at NITs 
including the appearance of unassigned bands in infrared (IR) spectra
 \cite{nidwir},  orders-of-magnitude 
enhancements of dielectric constants \cite{dielold,tokurajacs}, 
and  diffuse X-ray (DXR) signals \cite{dxdm,dxttf}.
 However, NIDWs were originally restricted \cite{nag3} to the
immediate vicinity of a discontinuous NIT, and the energy of domains
is in most cases too large to allow for appreciable  thermal 
population \cite{pes}. Here we demonstrate that 
soft modes and e-ph coupling provide alternative explanations
of all anomalies observed at NIT,
 without the restriction to a discontinuous NIT and 
without invoking either NIDWs or LR-CTs. 
 
\begin{figure}
\centering
\includegraphics* [scale=0.33]{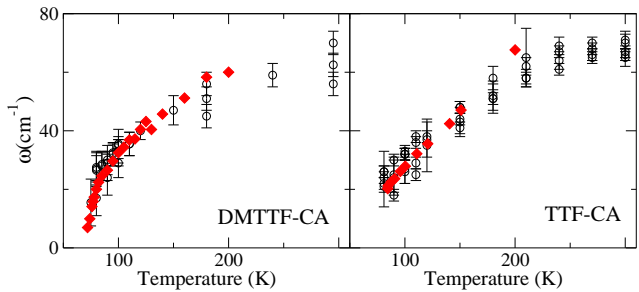}
\caption{Circles: the soft mode frequency from refs. \cite{softcpl,paolo}.
 Diamonds: the soft mode frequency (arbitrary units)  as
 inferred
on the basis of Eq. \ref{iofq}  from
  DXR data in Ref. \cite{dxttf,dxdm}.}
\label{speri}
\end{figure}

The first observation of a soft 
mode associated with NIT was in the mid-IR 
spectra of tetrathiafulvalene-chloranil (TTF-CA) \cite{softcpl}.
TTF-CA is the prototypical and best characterized CT salt, with a
discontinuous NIT at T$_c$ = 81 K from a regular neutral ($\rho \sim 0.3$) 
stack to a dimerized ionic ($\rho \sim 0.6$) stack. For T $>$ T$_c$
weak IR bands polarized along the stack axis appear
close to the frequency of Raman-active  A$_g$ molecular modes. These
bands, originally assigned to the NIDWs dynamics \cite{nidwir}, 
are combinations of the  A$_g$ modes with  the
IR-active dimerization phonon.
A careful analysis of IR 
and Raman spectra allowed for a reliable estimate of the 
frequency of the soft mode \cite{softcpl}. 
Similar results have been obtained
recently for dimethyl-TTF-CA (DMTTF-CA) \cite{paolo}, in which almost 
complete softening is observed. The circles in Fig. \ref{speri}
 show the
frequency of the soft mode in both crystals for  T $>$ T$_c$. 

The dielectric
peak at NIT was assigned to NIDW fluctuations \cite{dielold,tokurajacs}, 
 but its magnitude can be 
quantitatively modeled as due to the soft dimerization  mode that  
acquires a huge IR intensity near the transition \cite{delfreoprl,bpps}. 
DXR data \cite{dxdm,dxttf}  have been interpreted in terms of long
LR-CTs  in DMTTF-CA and short ones in TTF-CA: so far they offer the 
only unchallenged evidence for NIDWs. 

DXR scattering was widely used in the 80's to study soft modes 
that drive structural transitions, including the Peierls transition in
CT salts with a segregated stack \cite{pouget}.
 The Peierls transition in mixed (DA) stacks is however different:
 the soft mode or incipient soft
mode is an optical phonon instead of an acoustic phonon as in
segregated stacks. The sharp 
Lorentzian-shaped  DXR signal collected in one spatial direction 
implies a strong dispersion  of the relevant phonons and is unusual
for the optical branch of a molecular crystal. This was the basis for 
rejecting a soft mode interpretation in favor of LR-CTs \cite{dxttf}.
 But e-ph coupling is extremely effective in 1D and anomalous
 dispersion of optical phonons 
is well known in polyacetylene \cite{mele}.

Here we demonstrate that DXR data
can be quantitatively understood as due to the progressive softening
of the zone-center {\it optical} phonons  as the system is driven
towards the NIT. A sharp anomaly develops
in the optical phonon branch  that, although occurring in non-metallic systems, is analogous to the Kohn anomaly in 1D metals, and shares
the same origin, namely, e-ph coupling. Peierls transitions
were similarly  proposed for 1D metals and later
generalized to transitions driven
by e-ph coupling in insulators and spin chains.

The DXR intensity, $I$,  is related to the
frequency $\omega$ of the soft mode by a simple relation \cite{pouget}:
\begin{equation}
  I(q) \propto \frac{k_B T}{\omega^2(q)}
\label{iofq}
\end{equation}
where $q$ is the wavevector. Reliable estimates of
$\omega(0)$ were obtained from the analysis of combination bands in IR
spectra of  both TTF-CA and DMTTF-CA \cite{softcpl,paolo}. Since DXR 
scattering is in arbitrary units, Eq. \ref{iofq} yields the temperature
dependence of $\omega(0)$ to within a scale factor. The diamonds in 
Fig. \ref{speri} are scaled $\omega(0)$ from DXR. The striking coincidence 
of DXR and IR estimates of $\omega(0)$ for T $>$ T$_c$ in both salts 
provides strong motivation for a soft-mode analysis. But several
issues must be addressed for such an interpretation, starting with
the strong dispersion of the optical branch.

To confirm the soft-mode picture, we consider the dispersion of 
optical phonons in a 1D stack with equal spacing, harmonic force
constant $K$ for nearest-neighbor lattice displacements, molecular  masses 
for D and A, and linear e-ph coupling. For simplicity, we describe the
electronic structure of the mixed stack using a
spinless-fermion (SF) model,  an uncorrelated model
 \cite{pes} that can readily be solved for long chains.  
 The electronic Hamiltonian describes $N/2$
spinless fermions on  $N$ molecular sites with periodic 
boundary conditions, as follows:
\begin{equation}
 H_{SF}=\Gamma \sum_p(-1)^p \hat n_p-\sum_p t_p \hat b_p
\label{hsf}
\end{equation}
where $p$ runs on the sites, $\hat n_p$ counts the fermions on $p$ site,
$t_p$ is a transfer
integral, and  $\hat b_p=(a^{\dagger}_p a_{p+1}
+H.c.)$ is  the bond-order operator. In the SF model, a single
 fermion moves from an odd 
(D)  to an even (A) site to describe electron 
transfer between DA and  D$^+$A$^-$. 
$2\Gamma$ then represents the energy required to ionize a
 DA pair and implicitly includes all contributions from e-e
 interactions and from the coupling of electrons with molecular
 vibrations \cite{pes}. 
The hopping integrals are  modulated by lattice phonons:
$t_p=t_0+\alpha (x_p-x_{p+1})$ where $x_p$ is the displacement
of the $p$ site, and $t_0$
will be used as the energy unit.
 The e-ph coupling constant $\alpha$ enters the definition of the 
 vibrational relaxation energy:  $\ed=\alpha^2/K$. 
Aside from a spin factor, the model was discussed by Rice and
Mele \cite{rm}, and  at  $\Gamma$ = 0 it reduces to the SSH model
\cite{ssh}.
\begin{figure}
\centering
\includegraphics* [scale=0.4]{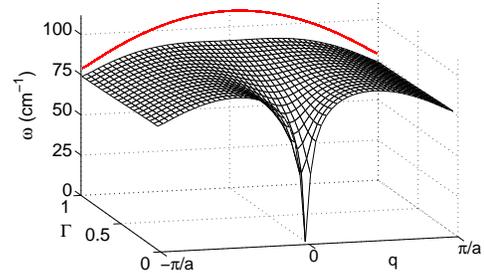}
\caption{The $\Gamma$-dependence of the optical dispersion curve
  $\omega(q)$, calculated for parameters relevant to DMTTF-CA, $\ed=0.4$ and
  $\omega_0=110$ cm$^{-1}$ on 902 sites. The continuous red line  is the
  dispersion curve for $\ed=0$. }
\label{disp}
\end{figure}

The  vibrational problem in the absence of e-ph coupling ($\ed=0$) 
describes a 1D harmonic chain with two masses. It has a well known
solution, and the red continuous line 
in Fig. \ref{disp} shows the dispersion curve for the optical branch,
$\omega_0(q)$, calculated for the molecular  masses
of DMTTF-CA and with  $K$ fixed as to get  $\omega_0(0)=110$ cm$^{-1}$.  
In the presence of e-ph coupling, $\ed >0$, 
new  harmonic force constants add to $K$, corresponding  to the second
derivatives of the ground state (GS) electronic energy $E$ vs the site 
displacements:
\begin{equation}
  \left(\frac{\partial ^2 E}{\partial x_p\partial
  x_{p+r}} \right)_{eq}=\alpha^2\left(\Pi_{r-1}  +
  \Pi_{r+1}- 2\Pi_{r}  \right)
 \label{polari}
\end{equation}
where  $\Pi_r=-(\partial^2E/\partial
t_p\partial t_{p+r})_{eq}$, the bond-bond polarizabilities, 
are {\it electronic}
quantities. In  the SF model:
\begin{equation}
  \Pi_r=2\sum_{k,l} \frac{\langle GS|\hat b_p|kl\rangle\langle kl|\hat b_{p+r}|GS\rangle}{\e_l-\e_k}
\label{bbp}
\end{equation}
where  $|kl\rangle$ is  the
excited state of $H_{SF}$ with a fermion  promoted from the 
$k$-th filled orbital to the $l$-th empty orbital, 
and $\e_{k}$ is the  $k$-orbital
energy. Much as in polyacetylene \cite{dellepiane},  the electronic
delocalization leads to  long-range
bond-bond polarizabilities,  and hence to long-range force constants
that  are responsible for an anomalous dispersion of the
optical phonon branch. The acoustic branch is only marginally affected
by e-ph coupling.

In momentum space,
the squared vibrational frequencies $\omega(q)^2$ are obtained, as
usual, by diagonalizing the  $2 \times 2$ 
force constant matrices, whose  elements  contain sums over 
$\Pi_{r}$ as well as $K$ and the molecular masses. 
The largest softening occurs at $q = 0$, where a simple expression
holds for the optical branch frequency \cite{pg86}:
\begin{equation}
  \omega(0)=\omega_0(0)\sqrt{1-\ed\chi(\Gamma)}
\end{equation}
where $\chi = -(\partial^2E(\Gamma,\delta)/\partial\delta^2)_0$,
 the curvature of the electronic 
GS energy along the dimerization coordinate, $\delta$, only depends
 on $\Gamma$. 
The  divergence of $\chi$ at $\Gamma = 0$ 
marks the unconditional Peierls instability of a half-filled 1D
 metal. For finite $\ed$ 
CT stacks are  stable against dimerization for $\Gamma >
 \Gamma_P$, where $\Gamma_P$ is defined by  
$\ed \chi(\Gamma_P) = 1$. 
Peierls transitions have been extensively  studied and there are  
analytical expressions for $\chi(\Gamma)$ 
for uncorrelated models \cite{handbook} and
numerical results for correlated models \cite{pg86,prb45}. 

Figure \ref{disp}
shows the evolution of $\omega(q)$ as function of $\Gamma >
\Gamma_P$.
For the chosen parameters, the Peierls transition occurs 
at $\Gamma_P =0.05$, where the softening is complete, $\omega(0) = 0$, 
 and a giant anomaly develops
in the dispersion of optical phonons. 
The softening of the optical branch follows 
directly from linear e-ph coupling. An important new 
result in Fig. \ref{disp}
is the explicit expression for the phonon dispersion as the dimerization
transition is approached. With $\omega(q)$
in hand, we can model the DXR profile 
$I(q)$ using Eq. \ref{iofq}.
  
\begin{figure}
\centering
\includegraphics* [scale=0.52]{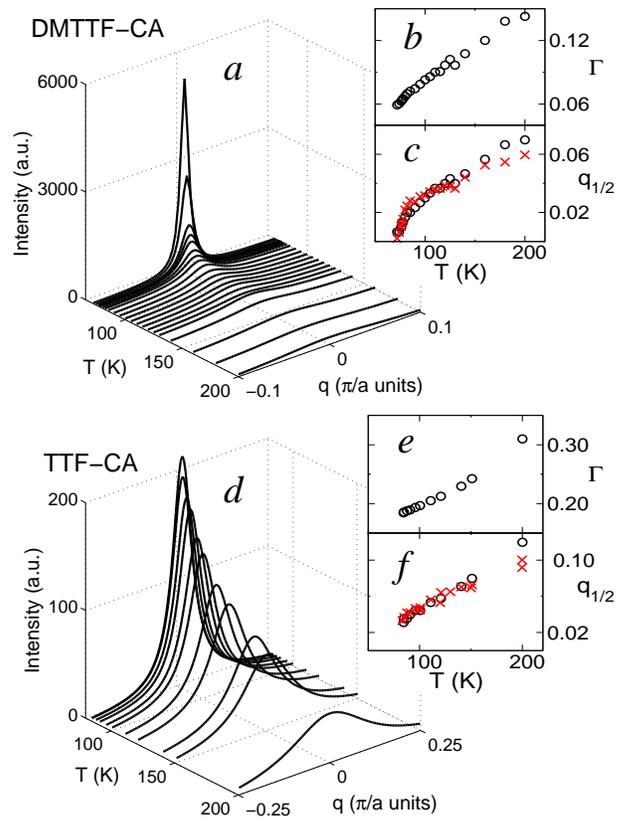}
\caption{Main panels ({\it a, d}): the T-evolution of the DXR signal as calculated
  in the SF model. Panel {\it a}: DMTTF-CA, $\ed=0.4$,
  $\omega_0(0)=110$ cm$^{-1}$;  panel {\it d}: TTF-CA, $\ed=0.55$,
  $\omega_0(0)=130$ cm$^{-1}$. Insets  {\it b} and {\it e} show the $\Gamma(T)$
  relation; {\it c} and {\it f} compare  the width of calculated DXR peaks
  (circles)  with the experimental values
  (crosses) from refs \cite{dxdm,dxttf}.}
\label{tutto}
\end{figure}

DXR profiles $I(q)$ are measured as a function of T, while the phonons 
in the SF model are computed as a function of $\Gamma$. The profile in 
either case is characterized by the peak height $I(0)$ and by the 
half-width at half-maximum, $q_{1/2}$.
 The calculated $I(q)$ in Fig. \ref{tutto}
 are  based on Eq. \ref{iofq}, with $\omega(0)$ taken from vibrational
 data, and the best-fit  parameters, 
$\e_{d}$ and $ \omega_0(0)$, in the 
caption. Insets {\it b} and {\it e}  
specify $\Gamma$(T) for DMTTF-CA and TTF-CA, respectively, 
 and an  almost linear relation is extracted in each case.
 The measured and calculated values of $q_{1/2}$ are seen in insets 
{\it c} and  {\it f}
 to be in excellent agreement, especially close to the transition. 
DXR profiles in mixed-stack CT salts close to the NIT
 can be quantitatively explained by the softening of the
dimerization mode and the related anomaly  in the optical phonon branch.

The sharp DXR peaks observed in DMTTF-CA were previously assigned
to the presence of long LR-CT excitations \cite{dxdm}, whereas the 
broader signals in TTF-CA were ascribed to shorter domains \cite{dxttf}. 
Here the difference between the two systems is quite naturally related 
to their different transitions. DMTTF-CA undergoes a continuous
(or almost continuous) dimerization transition \cite{paolo}: $\rho(T)$
is continuous, the phonon frequency $\omega(0)$ softens to zero (or
almost so), and the dip in the phonon dispersion 
 fully develops leading to sharp DXR peaks. 
On the other hand, only incipient soft-mode behavior is observed in
TTF-CA, whose discontinuous
NIT at 81 K interrupts the softening \cite{ricememorial}. As a result,
the dispersion anomaly does not fully develop and comparatively broad DXR peaks
are observed.

The non-interacting 
model, $H_{SF}$  in Eq.  \ref{hsf}, can be solved for very long 
chains, as needed to get
enough resolution in the dispersion curves. To show that 
the anomalous dispersion in the optical phonon branch survives
correlations, we now turn attention to the standard correlated model
for NIT, the modified Hubbard (MH) model  \cite{pes} with linear e-ph
coupling. MH describes $N$ electrons
on $N$ sites, and, as in SF,  $2\Gamma$ is the energy required to ionize a
DA pair. Correlations are taken into account by a restricted basis
that excludes $D^{2+}$ and $A^{2-}$ sites. Intersite e-e interactions
$V$ can be introduced in addition to linear e-ph coupling. 

\begin{figure}
\centering
\includegraphics* [scale=0.48]{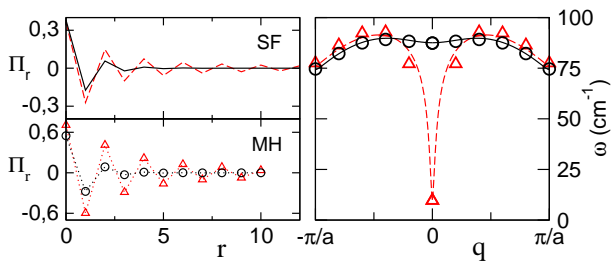}
\caption{Left panels: $r$-dependence of the bond-bond
 polarizabilities, calculated for
 $\rho$=0.46 (dashed line for SF, triangles for MH)
and 0.22 (continuous line for  SF, circles for MH). 
Right panel:  the dispersion curves calculated
 for the same $\rho$ as in the left panels, and parameters relevant to
 DMTTF-CA.
 Lines and symbols refer to SF ($N=902$) and  MH model ($N=20$).}
\label{bb}
\end{figure}

MH bond-bond polarizabilities $\Pi_r$ are calculated as numerical
derivatives of the exact GS energy of finite chains (up to 20 sites) 
with periodic boundary conditions.
We compare results for SF and MH chains with the same $\rho$, and in
the  left panels in Fig. \ref{bb} we 
show the $r$-dependence of the  $\Pi_r$ for two systems with
$\rho=0.22$ and 0.46, far and close to NIT, respectively. For both
systems, apart from 
a scale factor, the  $\Pi_r$ calculated in SF and MH models 
have similar behavior and  oscillate between positive
and negative values, much as found in
polyacetylene \cite{dellepiane}.   
The right panel of 
Fig. \ref{bb} compares the MH and SF phonon dispersion
curves calculated for the same $\rho$ values.  For the SF case we set  
the parameters relevant for DMTTF-CA. In the MH model, 
to account for the stronger e-ph coupling in correlated systems,  we 
rescale $\e_d$ to obtain the same  $\omega(0)$ as in  SF. 

Far from NIT, at $\rho=0.22$, the effect of e-ph 
coupling is mild: $\Pi_r$  decays rapidly 
with $r$, well within distances achievable
in a periodic 20 site chain.
As a result, the discrete MH points for $\omega(q)$
fall exactly on top of the SF curve in Fig. 4, right panel. On the
 contrary, near the transition 
($\rho=0.46$), $\Pi_r$ acquires a long-range tail 
that cannot be appreciated in a 20-site system. Nevertheless, the MH 
 points of the dispersion curves approximately fall on  the SF curve.
We conclude that the sharp anomaly in the optical phonon branch is a 
characteristic feature of NIT, and that its basic physics is well 
captured by models without e-e correlation.

In conclusion, we ascribe DXR scattering in DMTTF-CA 
and TTF-CA crystals on approaching the NIT 
 to the evolution of a Kohn-like anomaly in the 
optical phonon branch, related to the softening of the dimerization
phonon. The anomaly is due to 
amplified e-ph coupling near an electronic 
transition.  Our
interpretation, quantitatively consistent with vibrational frequencies
as extracted from mid-IR data, is supported by the calculation of
phonon dispersion curves 
in SF and MH models for electronic systems with 
linear (Peierls) e-ph coupling.
Linear e-ph coupling has enormous effects in systems with delocalized
electrons in 1D, as best demonstrated by polyacetylene
\cite{ssh,handbook}. In CT salts the amount of delocalization
increases when moving towards NIT from the N side, giving rise to
several anomalous features that have often been interpreted as
evidences for exotic excitations like NIDW or LR-CT. 
Such excitations are, however, only expected 
 in the immediate vicinity of a first-order transition \cite{nag3,pes}.
The soft mode interpretation of anomalous  mid-IR bands  
in CT salts appearing on  approaching NIT, as well as the huge peak
in the dielectric constant measured at NIT has already been discussed
\cite{softcpl,paolo,delfreoprl,bpps}.  
Here we showed that 
DXR scattering is another manifestation of e-ph 
coupling in 1D. 

\begin{acknowledgments}
Work in Parma supported
 by NE MAGMANET NMP3-CT2005-515767 and Italian MIUR through FIRB-RBNE01P4JF.
\end{acknowledgments}

\bibliography{soft}

\end{document}